\documentclass[twocolumn]{jpsj2}

\title{
Novel phase diagram of superconductor Na$_x$CoO$_2\cdot y$H$_2$O\\ in a 75 \% relative humidity
}

\author{%
Hiroto \textsc{Ohta}\thanks{E-mail address:
shioshio@kuchem.kyoto-u.ac.jp}, Chishiro \textsc{Michioka}, Yutaka \textsc{Itoh} and Kazuyoshi \textsc{Yoshimura}
}

\inst{%
Department of Chemistry, Graduate School of Science, Kyoto University,
Kyoto 606-8502
}

\recdate{\today}

\abst{%
We succeeded in synthesizing the powder samples of \textit{bi}layer-hydrate sodium cobalt oxide superconductors Na$_x$CoO$_2\cdot y$H$_2$O with $T_{\mathrm{c}}$ = 0 $\sim$ 4.6 K by systematically changing the keeping duration in a 75 \% relative humidity atmosphere after intercalation of water molecules. From the magnetic measurements, we found that the one-day duration sample does not show any superconductivity down to 1.8 K, and that the samples kept for 2 $\sim$ 7 days show superconductivity, in which $T_{\mathrm{c}}$ increases up to 4.6 K with increasing the duration. $T_{\mathrm{c}}$ and the superconducting volume fraction are almost invariant between 7 days and 1month duration. The $^{59}$Co NQR spectra indicate a systematic change in the local charge distribution on the CoO$_2$ plane with change in duration.
}

\kword{%
superconductivity,duration effect,sodium cobalt oxide,bilayer hydrate
}

\begin{document}
\maketitle

A remarkable outcome from the interplay between soft chemistry and solid state physics is the discovery of superconductivity of $bi$layer hydrate cobalt oxide Na$_x$CoO$_2\cdot y$H$_2$O.\cite{Takada_Nature} It arouses new interests in the geometric spin frustration effects on the superconductivity\cite{Baskaran,Khaliullin,PALee,PALee2,Kuroki}. From various viewpoints of chemistry, much effort has been devoted to control the superconducting transition temperature $T_{\mathrm{c}}$. At first, the amount of Na content was thought to change the carrier doping level of the CoO$_2$ plane and to lead to control $T_{\mathrm{c}}$ \cite{Schaak_Nature}. However, it turned out that such change of the Na content is not a unique parameter for $T_{\mathrm{c}}$ \cite{Milne_PD}. The chemical diversity of the $bi$layer hydrate superconductors was revealed by observations of ordering of Na occupied site \cite{HWZandbergen_sodiumorder} and the transformation of intercalated water molecules into oxonium ions at Na sites.\cite{Takada_oxonium}
Recently, controlling of the duration in a humidity was found to be effective to control $T_{\mathrm{c}}$ of three-layer hydrates.\cite{Foo_threeL} This study was a clue for us to control $T_{\mathrm{c}}$ of the $bi$layer hydrates. 

In this Letter, we report a successful synthesis of the $bi$layer hydrate Na$_x$CoO$_2\cdot y$H$_2$O with sequential $T_{\mathrm{c}}$'s from 0 to 4.6 K by controlling duration in a 75 \% relative humidity. We found the reproducible non-superconducting $bi$layer hydrate. The systematic change of $T_{\mathrm{c}}$ and also the superconducting volume fraction were confirmed by magnetic susceptibility measurements using a Superconducting Quantum Interference Device (SQUID) magnetometer.

\begin{figure}[h]
\begin{center}
\includegraphics[width=8cm,clip]{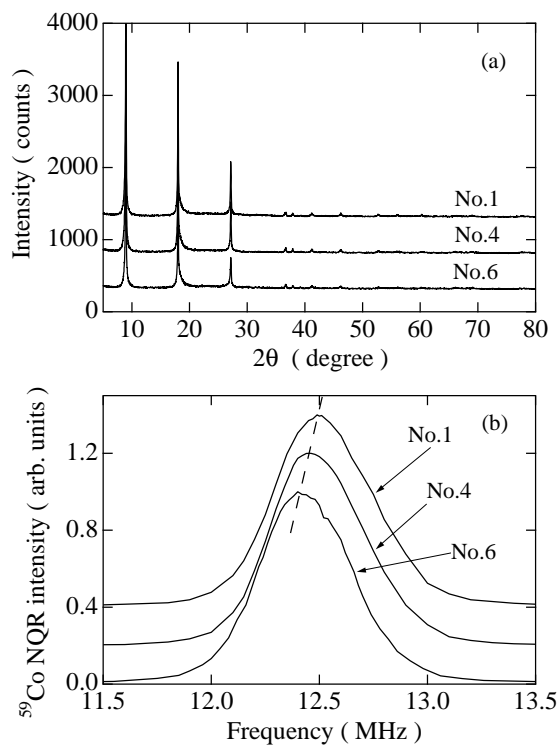}
\end{center}
\caption{(a)Typical patterns of powder X-ray diffraction for the samples No.1, No.4 and No.6 using Cu K$\alpha$ line at 20 $^{\circ}$C. Each sample is in a single phase of $bi$layer-hydrate. (b)Typical $^{59}$Co ( $I$=7/2 ) NQR spectra ( the highest transition lines of $I_{\mathrm{z}} = \pm5/2 \leftrightarrow \pm 7/2$ ) of the samples No.1, No.4 and No.6 at 8 K. The peak position shifts to lower frequency with increase in the duration. Dashed line is a guide to the eyes.
}
\label{fig:XRD}
\end{figure}

The parent compound Na$_{0.7}$CoO$_2$ was synthesized by so-called "rapid heating-up" method. \cite{Motohashi_rapid} Powders of Co$_3$O$_4$ ( 99.9 \% ) and Na$_2$CO$_3$ ( 99.99\% ) were ground in the mole ratio of 1 : 1.15, put in a pre-heated furnace at 750 $^{\circ}$C, and then kept for 20 hours. The resultant preproducts were again well ground, heated to 800 $^{\circ}$C for 8 hours, and kept for 16 hours with flowing oxygen gas.  Analysis of powder X-ray diffraction patterns indicates a single phase of Na$_{0.7}$CoO$_2$.

The powder of Na$_{0.7}$CoO$_2$ was immersed in 6 M Br$_2$/CH$_3$CN solution for one day to deintercalate Na$^+$ ions. The amount of the bromine was 10 times larger than the necessary amount for reduction of Na content from 0.7 to 0.4. After filteration, the powder was vacuumed to completely remove bromine and acetonitrile. The well dried powder of 1 g was immersed and stirred in distilled water of 300 ml for 16 hours to intercalate H$_2$O molecules, then the powder was filtered by filteration of absorption method. At this stage, the obtained powder was characterized to be $bi$layer hydrated Na$_x$CoO$_2\cdot y$H$_2$O. 

The obtained Na$_x$CoO$_2\cdot y$H$_2$O was kept in a 75 \% humidified chamber, and the power of 0.1 g was picked out from the chamber day by day, quickly preserved in a freezer at -10 $^{\circ}$C. As shown below, the physical properties of Na$_x$CoO$_2\cdot y$H$_2$O depend on this duration (keeping time in the humidified chamber). The cryopreservation in the freezer plays a key role in quenching further change. We prepared eight samples, which duration were 1 to 7 days and 1 month. Hereafter, the sample with duration of \mbox{$n$(= 1, 2, $\dots$ 7, 30 )} days is denoted by No.$n$. 

\begin{figure}[t]
\begin{center}
\includegraphics[width=8cm,clip]{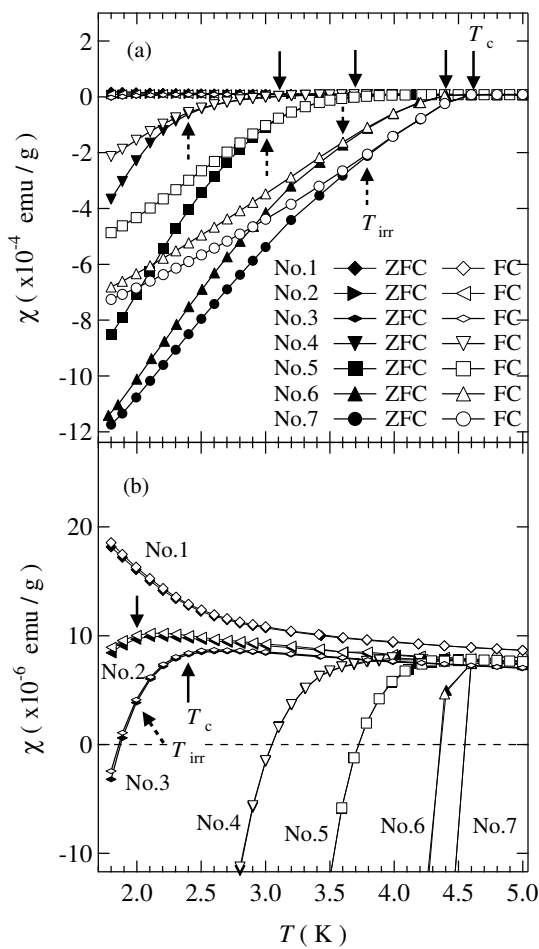}
\end{center}
\caption{(a) Temperature dependence of $\chi$ measured at 20 Oe for the samples of No. 1 $\sim$ No. 7. The data ( closed and open symbols ) were measured after ZFC and after FC, respectively. Solid and dashed arrows indicate $T_{\mathrm{c}}$ and $T_{\mathrm{irr}}$, respectively. (b) Extended plots of Fig. (a).}
\label{fig:chiT20Oe}
\end{figure}

Figure \ref{fig:XRD}(a) shows the powder X-ray diffraction patterns of samples No.1, No.4 and No.6 using the Cu K$\alpha$ line at 20 $^{\circ}$C. From the patterns, each sample was found to be in a single phase of $bi$layer-hydrate.  Lattice constants were estimated by using diffractions \mbox{(002)}, \mbox{(004)}, \mbox{(006)}, \mbox{($\bar{1}$12)} and \mbox{(01$\bar{4}$)}. The values of lattice constants \mbox{( $a$} $\sim$ 2.824 {\AA} and $c \sim$ 19.705 {\AA} ) did not show any obvious change against the increase in the keeping duration.

Figure \ref{fig:XRD}(b) shows the $^{59}$Co ( $I$ = 7/2 ) nuclear quadrupole resonance (NQR) spectra of the samples No.1, No.4 and No.6 at 8 K. The peaks correspond to the $I_{\mathrm{z}} = \pm 5/2 \leftrightarrow \pm 7/2$ transition lines. In contrast to the X-ray diffraction results, an appreciable change was observed in the $^{59}$Co NQR spectra, whose peak frequency gradually decreases with increase in the keeping duration. Dashed line in the figure guides the shift of the peak. Our samples show systematic increase in $T_{\mathrm{c}}$ with increasing the duration as shown below. Thus, the duration effect results in no appreciable change in the lattice constants but a obvious change in the local charge distribution probed by $^{59}$Co NQR spectra.

\begin{figure}[t]
\begin{center}
\includegraphics[width=7cm,clip]{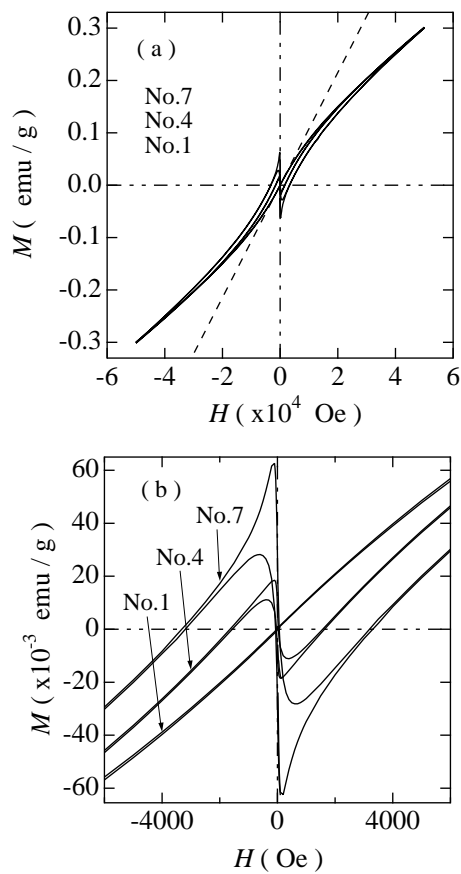}
\end{center}
\caption{(a)Magnetic field dependence of magnetization ($M$-$H$ curves ) of No.1, No.4 and No.7 measured at 1.8 K. The dashed line has the initial slope of the No. 1 sample near a zero magnetic field. (b) Extended plots of Fig. (a). Superconducting hysteresis develops with increasing the duration.}
\label{fig:MH}
\end{figure}

Figure \ref{fig:chiT20Oe} shows the temperature dependence of the magnetic susceptibility $\chi$ for the samples No. 1 $\sim$ 7 at 20 Oe and at 1.8 $\sim$ 5 K. The data ( closed and opened symbols ) indicate the $\chi$ measured after zero field cooling (ZFC) and field cooling (FC), respectively. All the high temperature $\chi$ above 5 K (not shown here) show similar behavior to the previously reported one\cite{Takada_Nature}. Solid arrows indicate superconducting transition temperatures $T_{\mathrm{c}}$'s, which are defined by the onset of superconductivity. The sample \mbox{No. 1} does not show superconductivity down to 1.8 K. $T_{\mathrm{c}}$ of No. 2 $\sim$ 7 increases with increasing duration. 
Dashed arrows indicate the irreversibility temperatures $T_{\mathrm{irr}}$'s, which show similar duration dependence to $T_{\mathrm{c}}$. 

\begin{figure}[t]
\begin{center}
\includegraphics[width=7cm,clip]{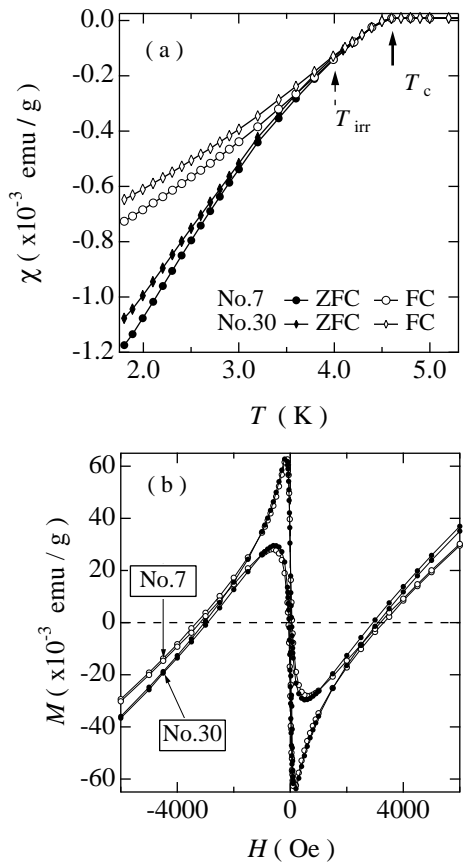}
\end{center}
\caption{(a) Temperature dependence of the magnetic susceptibility $\chi$ of the samples No.7 and No.30. The data denoted by closed symbols are measured after ZFC, and the data denoted by opened symbols are after FC. Solid arrow and dashed arrow indicate $T_{\mathrm{c}}$ and $T_{\mathrm{irr}}$, respectively. (b) $M$-$H$ curves of the samples No. 7 and No. 30 measured at 1.8 K. The data of No.7 and No.30 are shown by solid line with opened circles and solid line with closed circles, respectively. }
\label{fig:mag7vs30}
\end{figure}

\begin{figure}[h]
\begin{center}
\includegraphics[width=8cm,clip]{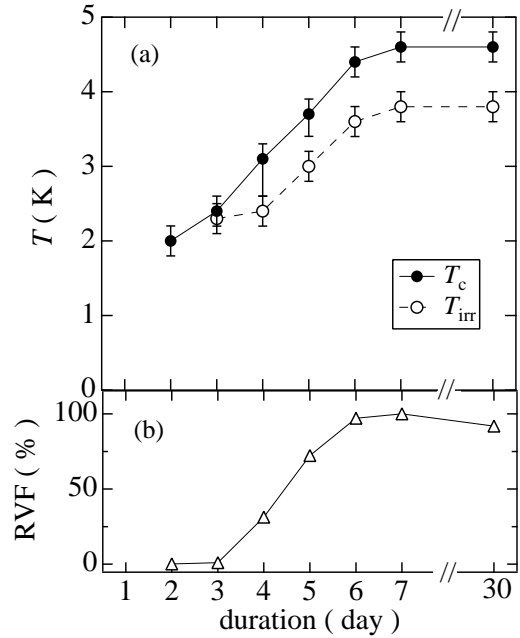}
\end{center}
\caption{$T_{\mathrm{c}}$, $T_{\mathrm{irr}}$ and relative volume fraction ( RVF ) plotted against the duration. $T_{\mathrm{c}}$ and $T_{\mathrm{irr}}$ increase with duration and saturate after 7 days. The RVF shows similar behavior to $T_{\mathrm{c}}$.}
\label{fig:Value_estim}
\end{figure}

Figure \ref{fig:MH} shows the magnetic field dependence of magnetization of the samples No. 1, 4, and 7 at a magnetic field of $H \le$ 5 T and at 1.8 K. The slope of the dashed line in Fig. \ref{fig:MH}(a) is the same as an initial slope near zero magnetic field for the sample No. 1, which does not show superconductivity. The nonlinear high field magnetization of the sample No. 1 in Fig. \ref{fig:MH}(a) suggests ferromagnetic correlation. The another samples show similar behavior at a high field. As shown in Fig. \ref{fig:MH} (b), the superconducting hysteresis develops from narrow to wide hysteresis loops with increasing the duration, being consistent with the behavior of the temperature dependence of $\chi$. 

Figure \ref{fig:mag7vs30} (a) and (b) show the temperature dependence of $\chi$ and the magnetization curves at 1.8 K for the samples No.7 and No.30. Both physical properties show similar in both samples, indicating that the duration effect is small between 7 and 30 days. Especially, $T_{\mathrm{c}}$ and $T_{\mathrm{irr}}$ are invariant. The irreversibility field $H_{\mathrm{irr}} \sim$ 2.5 T of No.30 at 1.8 K, however,  is about 6 times larger than $H_{\mathrm{irr}} \sim$ 0.4 T of No. 7. This result indicates that the vortex pinning number and the pinning force increase from 7 days to 30 days. The upper critical field $H_{\mathrm{c2}}$ may increase from 7 days to 30 days.

Figure \ref{fig:Value_estim} shows the duration dependences of  $T_{\mathrm{c}}$, $T_{\mathrm{irr}}$ and relative superconducting volume fraction. The relative volume fraction is normalized by the value of No. 7 sample. All these values for superconductivity increase as duration increases and almost saturate at 7 days. From the NMR and NQR results, the duration effect was found to change the local charge distribution around the in-plane Co nuclei and the inter-plane correlation.\cite{Ohta_NMR} The present novel phase diagram of Fig. \ref{fig:Value_estim} indicates that the superconductivity happens through the strong correlation with a delicate difference in the amount of intercalated water molecules.

In this study, we found that the non-superconducting sample can be obtained by short duration treatment after intercalation of water molecules. By a change of the condition, we have already confirmed that a sample does not show superconductivity even at 7 days duration. The duration effect depends on the initial condition of the parent powders ( Na content and the size of the \mbox{particle )} and the condition of the humidified chamber \mbox{( humidity} and temperature ). Now the detailed studies of multiple-phase diagram on duration effect by controlling the initial condition and the atmosphere in the chamber is in progress. 

We thank Dr. M. Kato for his helpful discussions.
This work was supported in part by Grants-in Aid for Scientific Research
from Japan Society for the Promotion of Science (16076210).

\end{document}